# The chemical composition of the Galileian satellites

Vladan ^elebonovi}
Institute of Physics, Pregrevica 118,11080 Zemun- Beograd, Yugoslavia

celebonovic@ exp. phy. bg. ac. yu
vcelebonovic@ sezampro. yu

Abstract: Using the semiclassical theory of the behaviour of materials under high pressure proposed by P.Savi} and R.Ka{anin the mean molecular masses of the Galileian satellites are determined. The numerical values are fitted by plausible combinations of chemical elements, and possible cosmogonical explanations of the results are discussed to some extent.

## Introduction

The chemical composition of celestical bodies can be determined either by remote spectroscopy or within theories attempting to explain their origin and internal structure. According to recent reviews ( Encrenaz,1984; Stevenson,1985;Zellner et al.,1985 ) , our knowledge of the chemical composition of the smaller bodies of the solar system is rather limited.

## Results

The aim of this letter is to determine theoretically the chemical composition of the Galileian satellites. All the calculations were performed within the semiclassical theory of the behaviour of materials under high pressure, proposed by P.Savi} and R.Ka{anin ( Savi} and Ka{anin, 1962/65 ; ^elebonovi},1986; Savi} and Teleki,1986 and references given there ). This study has been undertaken with the aim of testing the applicability of the theory to planetary satellites and, in case of satisfactory agreement, extending it to the determination of the composition of the asteroids.

As input data we have used the masses and radii of the satellites (Masson,1984 ).The results are presented in the following table, in which A denotes the mean molecular mass of the mixtures which can approximate the composition of different satellites

| SATELLITE | A | MIXTURE |
|---|---|---|
| Io | 70 | $22\% \, FeSiO_3 + 20\% \, FeS +$ $+ 28\% \, SO_2 + 30\% \, N_2H_4H_2O$ |
| Europa | 71 | |
| Ganymede | 18 | $85\% \, SiO_2 + 7\% \, H_2O +$ $+ 8\% \, H_2$ |
| Callisto | 19 | |

Discussion and conclusion

A detailed comparison of our results with observational data (Dollfus,1975 ; Masson,1984 and many other references ) reveals that the mixtures by which we have described the composition of the satellites contain most of the actually observed elements.

As for the origin of the distribution of values of A shown in the table,it can be explained by the depletition of the inner parts of the circum - jovian accretion disk in light elements, due to the excess luminosity of proto jupiter (Graboske et.al.,1975;Masson, 1984 ).

In conclusion, it has been shown that this theory can be used in studies of planetary satellites, and that it seems reasonable to attempt using it in studies of asteroids.

Note ( added July 20[th],1998 )

This paper was written in 1987., and presented at the II Workshop "Astrophysics in Yugoslavia", held on September 8 - 10,1987. at the Astronomical Observatory in Beograd .It was published in the Book of Abstracts ( ed. by M.S. Dimitrijevi} ),p.41. The results are in good agreement with the observations made by the "Galileo" probe nearly ten years later ( astro-ph/ 9607073 ). For a modern review of the theory see astro-ph/9803213 and references given there.

References

^elebonovi},V.: 1986,Earth,Moon and Planets,**34** ,59.
Dollfus, A.: 1975,Icarus, **25** , 416.
Encrenaz, Th.: 1984,Space Sci.Rev.,**38** , 35.
Graboske, H. C., Pollack, J. B., Grossman, A. S. and Olness, R. J.: 1975, Astrophys.J.,**199**,265.
Masson,P.: 1984,Space Sci.Rev.,**38** ,281.
Savi},P. and Ka{anin,R.: 1962/65,The Behaviour of Materials Under High Pressure, I-IV,Ed.Serbian Academy of Sciences and Arts,Beograd.
Savi},P. and Teleki,G.: 1986,Earth,Moon and Planets, **36** ,139.
Stevenson,D.J.: 1985,Icarus, **62** ,4.
Zellner,B.,Tholen,D.J. and Tedesco,E.F.: 1985,Icarus,**61**, 355.